\documentclass{article}
\usepackage{spconf,amsmath,amsfonts,graphicx}
\usepackage{booktabs}

\title{Enhanced Hierarchical Music Structure Annotations via Feature Level Similarity Fusion}
\twoauthors
  {Christopher J. Tralie \sthanks{C. Tralie was supported by an NSF RTG grant NSF-DMS 1045133.}}
	{Duke University\\
	Department of Mathematics\\
	Durham, NC, USA}
  {Brian McFee}
	{New York University\\
	Music and Audio Research Lab \& Center for Data Science\\
	New York, NY, USA}

\begin{document}
%
\maketitle
\begin{abstract}


 We describe a novel pipeline to automatically discover hierarchies of repeated sections in musical audio. The proposed method uses similarity network fusion (SNF) to combine different frame-level features into clean affinity matrices, which are then used as input to spectral clustering. While prior spectral clustering approaches to music structure analysis have pre-processed affinity matrices with heuristics specifically designed for this task, we show that the SNF approach directly yields segmentations which agree better with human annotators, as measured by the ``L-measure'' metric for hierarchical annotations. Furthermore, the SNF approach immediately supports arbitrarily many input features, allowing us to simultaneously discover structure encoded in timbral, harmonic, and rhythmic representations without any changes to the base algorithm.

\end{abstract}
\section{Introduction}\label{sec:intro}

Music has structure along many axes, such as timbre, melody, harmony, rhythm, etc.  Since most methods for automatic music structure segmentation are tuned to find a particular kind of structure, extending them support other types is usually quite difficult.  Hence, we would like to explore how to leverage multiple representations of the same audio to efficiently discover musical structure.

\subsection{Our contributions}

In this work, we show how to use similarity network fusion (\cite{wang2012unsupervised,wang2014similarity}, Section~\ref{sec:SNF}) for musical structure analysis.  In particular, our proposed method integrates disparate representations of timbre, harmony, and rhythm (Section~\ref{sec:features}) to produce a unified structure representation.  We then couple this method with spectral clustering (Section~\ref{sec:spectral}) to produce multi-level structure analyses, and we evaluate the system for its ability to recover annotated structure in a diverse music collection (Section~\ref{sec:evaluation}).  Our evaluation includes multiple reference annotations for each track, accounting for subjectivity and diversity of opinion. Overall, we find that our proposed method is more robust than prior work, and gets closer to human-level agreement than prior work.  L-recall (Section~\ref{sec:lmeasures}) is particularly strong with our technique, with a mean of 0.658 compared to the human inter-annotator recall mean of 0.664.

\subsection{Related work}

Similarity network fusion (SNF) is a joint random walk technique that was devised to leverage the strengths of different hand-designed similarity measures for shape classification 2D contours in images~\cite{wang2012unsupervised}.  It has since been used in such tasks as cancer phenotype discrimination~\cite{wang2014similarity}, image retrieval~\cite{chen2018ci}, and drug taxonomy~\cite{chen2018ci}.  SNF was introduced to the music information retrieval community by the authors of \cite{Chen2017CSFusion} to leverage different cross-similarity alignment scores in automatic cover song identification.  As in the original application, they use SNF {\em at the object (song) level}.  By contrast, it was shown in~\cite{tralie2017cover} that using SNF at the {\em feature level} (i.e., beat-synchronous HPCP and MFCC) can improve cross-similarity matrices between pairs of covers without the need for a network of song-level similarity measures.  A precursor to our work used SNF on frame-level features within a song to improve self-similarity matrices for visualization~\cite{tralie2018graphditty}.

As for music structure analysis, the present work builds directly upon the Laplacian spectral decomposition (LSD) method~\cite{mcfee2014spectral}.
This method operates by carefully constructing a graph which encodes short-term timbral continuity along with long-term harmonic repetition, and then partitions the graph at multiple scales to recover multi-level segmentations. While this can be effective, the graph construction depends heavily upon the choice of input features, and the resulting method can be somewhat brittle in practice. The method we propose here, in contrast, supports the fusion of arbitrarily many input representations, which facilitates the discovery of both long- and short-range structure along many different musical dimensions, including timbre, harmony, and rhythm.

\section{Methods}

\subsection{Fusion}\label{sec:SNF}

\begin{figure}
    \centering
    \includegraphics[width=0.95\columnwidth]{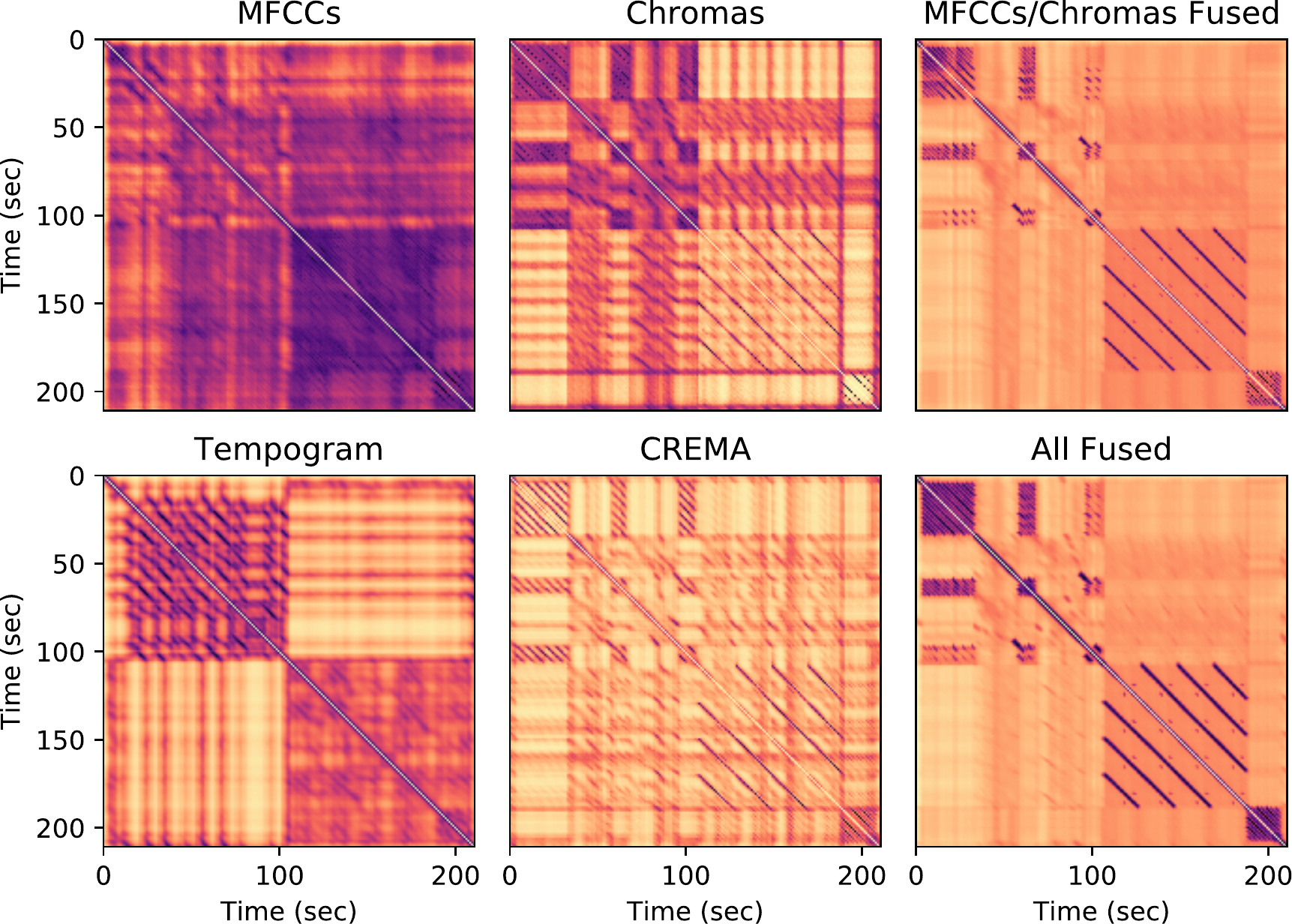}
    \caption{Applying SNF on the song ``Tango Apasionado'' by Astor Piazzolla (936 in the SALAMI dataset~\cite{smith2011design}).  Affinity matrices are shown before and after fusion.}\label{fig:SSMs}
\end{figure}

\label{sec:spectral}
\begin{figure}
    \centering
    \includegraphics[width=0.95\columnwidth]{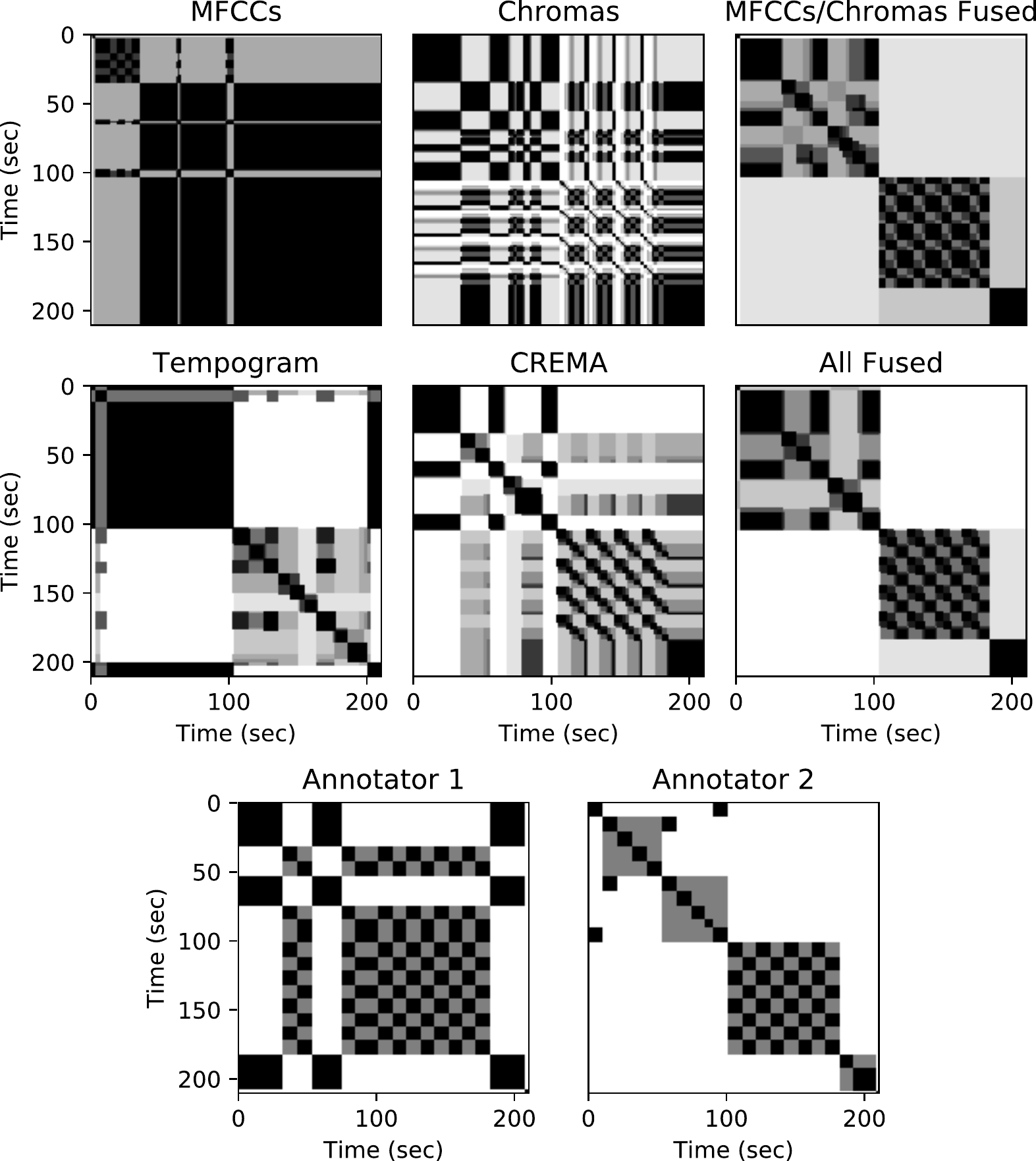}
    \caption{Applying spectral clustering to the affinity matrices in Figure~\ref{fig:SSMs}.  Meet matrices \cite{mcfee2017evaluating} are shown before and after fusion, in addition to meet matrices from human annotators.}\label{fig:MeetMatrices}
\end{figure}

We now provide details of frame-level similarity network fusion.  Given $F$ sets of features which are each computed at the same $N$ time intervals each corresponding to a stack-delayed sequence of frames (Section~\ref{sec:features}), we first compute the corresponding $F$ $N \times N$ self-similarity matrices (SSMs) $\{D^{f}\}_{f=1}^F$ via feature specific distances (Section~\ref{sec:features}).  Then, we convert each SSM to an ``affinity matrix'' $W^{f}_{ij} = \exp \left( -(D^{f}_{ij} / \sigma^{f}_{ij})^2 \right)$ with a pairwise autotuned time-dependent spatial bandwidth $\sigma^{f}_{ij}$, so that, as prescribed by~\cite{wang2012unsupervised} and~\cite{wang2014similarity}
\begin{equation}
    \sigma^{f}_{ij} = \frac{1}{6} \left( \frac{1}{\kappa} \left( \sum_{k \in N^{f}_\kappa(i)} D^{f}_{ik} +   \sum_{k \in N^{f}_\kappa(j)} D^{f}_{kj} \right) + D^{f}_{ij} \right)
\end{equation}
where $\kappa$ the number of nearest neighbors which is fixed a priori (we will explore the effect of $\kappa$ in Section~\ref{sec:results}), and $N^{f}_\kappa(i)$ are the indices of the $\kappa$ nearest neighbors of $i$, as measured by $D^{f}$.  SNF then defines two additional normalized versions $P^{f}$ and $S^{f}$ of each affinity matrix as follows
\begin{equation}
    P^{f}_{ij} = \left\{ \begin{array}{cc} W^{f}_{ij} / (2 \sum_{k \neq i} W^{f}_{ik}) & j \neq i \\ 1/2 & j = i \end{array} \right\}
\end{equation}
\begin{equation}
    S^{f}_{ij} = \left\{ \begin{array}{cc} W^{f}_{ij} / (2 \sum_{k \in N^{f}_\kappa(i)} W^{f}_{ik}) & j \in N^{f}_\kappa(i) \\ 0 &  \text{otherwise} \end{array} \right\}
\end{equation}
In other words, each $P^{f}$ can be interpreted as a doubly-stochastic transition probability matrix associated to $W^{f}$, and $S^{f}$ is the nearest neighbor thresholded version of $P^{f}$.  Given these matrices, SNF proceeds with the following iterations
\begin{equation}
    P^{f}_t = S^{f} \times \left( \frac{\sum_{k \neq f} P^{k}_{t-1}}{F-1}  \right) \times (S^{f})^{\mathsf{T}}
\end{equation}
for $T$ iterations $t = 1, 2, \dots, T$, cycling through $f = 1, 2, \dots, F$ at each iteration, and with $P^{f}_1 = P^{f}$ and $S^{f}$ fixed.  The final fused affinity is then taken to be $\frac{1}{F} \sum_{f = 1}^{F} (P^{f}_T)$. Intuitively, each iteration for feature type $f$ performs a random walk using neighbors of $f$ but probabilities from the other feature types averaged together, thereby fusing information from all features.  These iterations have been shown to converge quickly in practice~\cite{wang2012unsupervised,wang2014similarity}, and we find that $T = 10$ suffices.

\subsection{Spectral clustering}

Once we have clean affinity matrices, we can extract segments from them via spectral clustering~\cite{von2007tutorial}. Spectral clustering refers to a family of methods for partitioning graphs based on the characteristics of the eigenvector decomposition of their Laplacian matrix.
In this work, we use the \emph{random-walk normalized Laplacian} formulation.
Given a symmetric graph affinity matrix $A \in \mathbb{R}_+^{N\times N}$, the normalized Laplacian is defined as
\begin{equation}
L := I - \Delta^{-1}A,
\end{equation}
where $\Delta = \text{diag}(A\mathbf{1})$ is the diagonal \emph{degree} matrix of $A$.

$L$ is positive semi-definite, and the eigenvectors associated with the smallest eigenvalues encode the large-scale structure of the graph.
Let $L$ have eigenvector decomposition $L = \sum_i \lambda_i v_i v_i^\mathsf{T}$ with $\lambda_i$ in ascending order.
Spectral clustering proceeds by using the first $k$ eigenvectors as $V_k = [v_0, v_1, \dots, v_{k-1}] \in \mathbb{R}^{N \times k}$ as a $k$-dimensional feature representation of the nodes of the graph, which is then given as input to a $k$-means clustering algorithm (we use sklearn for $k$-means~\cite{pedregosa2011scikit}).
The resulting cluster assignments provide a partition of the nodes of the graph into $k$ disjoint subsets.

This general idea was previously applied to multi-level music segmentation~\cite{mcfee2014spectral} by iterating over multiple values of $k$: small values of $k$ produce few segment \emph{types}, though potentially many individual \emph{segments} of each type.
For each value of $k$, segment boundaries are inferred by finding the nodes $(n, n+1)$ (corresponding to time or beat indices) which receive distinct cluster assignments.  In this work, we create a hierarchy of segment labels by varying $k$ from 2 to 10. 

Figure~\ref{fig:MeetMatrices} shows an example of ``meet matrices''~\cite{mcfee2017evaluating} on the results of spectral clustering on the affinity matrices from Figure~\ref{fig:SSMs}.  Darker pixels in these matrices correspond to regions which are more consistently labeled across different levels in the label hierarchy. As in~\cite{grohganz2013converting} and~\cite{mcfee2014spectral}, we observe that tight diagonals in the SSMs are expanded as blocks in the annotations.  Hence, since SNF enhances diagonals in this example (Figure~\ref{fig:SSMs}), it leads to cleaner block structures.

\subsection{Features}\label{sec:features}

We evaluate the proposed fusion clustering method using four different audio representations, meant to encode various aspects of timbre, harmony, and rhythm.
Features are computed with librosa 0.6.2~\cite{brian_mcfee_2018_1342708}, and sampled at a framerate of 23.2~ms.

As a coarse timbre descriptor, we use 20 mel frequency cepstral coefficients (MFCCs), derived from a 128-dimensional mel spectrum covering 0--11025Hz.  We apply an exponential lifter $\hat{x_c} = (c^{0.6}) x_c$, to each coefficient $x_c, c=1, 2, \hdots, 20$.

To encode harmonic content, we use chroma derived from a constant-Q spectrogram of 36 bins per octave.
Chroma features capture harmonic content by aggregating pitch class energy across octaves, and can therefore be sensitive to overtones and transients.
To capture longer-term harmonic stability, we introduce a second set of features derived from the CREMA chord estimation model~\cite{mcfee2017structured}.
This model uses  convolutional-recurrent neural network for large-vocabulary chord recognition, and as a byproduct, produces conditional likelihood of each pitch class being active at each frame.
While these features can be interpreted as chroma-like, the recurrent aspect of the model tends to enforce local consistency while suppressing transients and passing tones.\footnote{CREMA features are produced at a framerate of 44100/4096=10.7Hz, and up-sampled by nearest-neighbor interpolation.}

Finally, rhythmic content is encoded by a tempogram derived from the local auto-correlation of the onset strength envelope~\cite{grosche2011tempogram}.
The onset strength envelope is calculated with a SuperFlux~\cite{bock2013maximum} local max filter of 5 bins on the previous frame, which suppresses vibrato while preserving attack transients~\cite{bock2013maximum}.
Tempogram auto-correlations are estimated over a window of 384 frames ($\sim$ 9 seconds) and peak-normalized.


After the initial computation, all features are averaged within non-overlapping chunks of 10 windows, slowing the framerate down to 0.232 seconds.  Unlike other works, we keep this constant across all songs, rather than using beat-synchronous sampling, which is can be brittle on certain genres.  Next, to promote temporal continuity when comparing windows, we stack delay overlapping blocks of 20 windows for each feature, as in~\cite{serra2009cross}.  Hence, each block spans roughly 4.64 seconds.  We then compute Euclidean SSMs between the MFCC and tempogram blocks, and we compute SSMs based on the cosine distance between the Chroma and CREMA blocks.  For a marginal improvement, we can enhance temporal continuity in a manner similar to~\cite{mcfee2014spectral} by performing a 9-tap median filter on each diagonal of the affinity matrices for each feature before applying SNF.

\section{Evaluation}\label{sec:evaluation}

Below, we describe the data and summary statistics we use.  As a baseline algorithm, we compare to LSD~\cite{mcfee2014spectral} as implemented in MSAF~\cite{nieto2016systematic}.

\subsection{Data}

We use the SALAMI dataset~\cite{smith2011design} with multilevel annotation corrections~\cite{mcfee2017evaluating} to quantiatively evaluate our algorithm.  This dataset consists of 1,359 tracks across a wide variety of genres which each have at least one annotator who has marked ``coarse'' and ``fine'' segments.  In our work, we focus on a subset of 884 songs which have two distinct annotators, so that we can compare to a human-level annotator agreement.

\subsection{Evaluation criteria}

Numerous methods have been proposed to evaluate the accuracy of musical structure estimation systems.
For most choices of evaluation criteria (e.g., segment boundary accuracy or segment labeling), there are two critical sources of variation which must be accounted for: ambiguity in structural depth, and subjectivity across reference annotators.\footnote{All evaluations are implemented using \texttt{mir\_eval} 0.5~\cite{raffel2014mireval}}

\subsubsection{L-measures for hierarchical structure}\label{sec:lmeasures}

The L-measure~\cite{mcfee2017evaluating} was proposed as a generalization of pairwise frame classification~\cite{levy2008structural} to support comparison between multi-level time-series segmentations, which we briefly summarize here.
Multi-level segmentations are assumed to be provided as a sequence of collections of labeled intervals $H = (\Pi_0, \Pi_1, \dots)$, where each $\Pi_i$ partitions the input signal in time, and the sequence is ordered from \emph{coarse} to \emph{fine}.
Typically, $\Pi_0$ is a single interval which spans the entirety of the input, and subsequent $\Pi_i$ provide refinements into collections of smaller segments.

If $\Pi_i(t)$ denotes the label of the interval containing time $t$ at the $i^\text{th}$ level of the segmentation, then a similarity between instants $(t, u)$ can be derived from the maximum $i$ such that $\Pi_i(t) = \Pi_i(u)$.
This pairwise similarity function gives rise to a partial ordering over pairs of time instants, which can be summarized by the set of all triples $(t, u, v)$ such that $(t,u)$ are more similar than $(t, v)$.
Given two multi-level segmentations $H^R$ (the reference) and $H^E$ (the estimate), the L-precision (L-recall) is defined as the fraction of such triples in the estimate (reference) also found in the reference (estimate).
The L-measure is defined as the harmonic mean of L-precision and L-recall.

As demonstrated in prior work, the L-measures facilitate holistic comparison between multi-level segmentations of differing depths, and are robust to level-alignment errors~\cite{mcfee2017evaluating}.
These properties make them well-suited to evaluating the ability of the proposed fusion method to capture multiple forms of structure in music.
Note that because the estimators under comparison in this work all produce annotations of greater depth than the reference annotations---which all have two non-trivial levels---the precision scores may not be reliable.
We therefore focus our evaluation on L-recall, which measures how much structure in the reference annotation was identified in the estimate.
However, for completeness, we provide a full report of L-precision, L-recall, and L-measure.

\subsubsection{Inter-annotator agreement comparison}

Most evaluations of music structure analysis systems assume a single \emph{ground truth} reference annotation for each track, compare the system's estimate to that reference, and summarize the distribution of evaluation scores over all tracks, e.g., by reporting the mean score.
However, recent work has shown that multiple annotators often exhibit divergence of interpretation of musical structure, and this variation should be taken into account when evaluating systems.

We follow the design of~\cite{mcfee2017evaluating}, and compare the distributions of L-measures when comparing an estimator to \emph{multiple} reference annotations to the distribution of scores arising from comparing the annotators to \emph{each other}.
Using the subset of the SALAMI collection for which we have multiple reference annotations (each containing multi-level segmentations), we compute the L-measure scores for each pair of annotations for each track, producing a sample of scores $p_a$.
For each estimator $e$, we then compare each estimated structure to all annotations, which results in a second sample of scores $p_e$.
The collections $p_a$ and $p_e$ are then compared using the two-sample Kolmogorov-Smirnov test statistic ($K$), which measures the maximum absolute difference between their (discrete) cumulative distribution functions: small values indicate similar distributions.
This comparison measures the performance of the estimator relative to inter-annotator disagreement.
For completeness, we also report the mean L-measure scores (across all tracks and annotators) to provide an absolute measure.

\subsection{Results}\label{sec:results}
\begin{figure}
    \centering
    \includegraphics[width=\columnwidth]{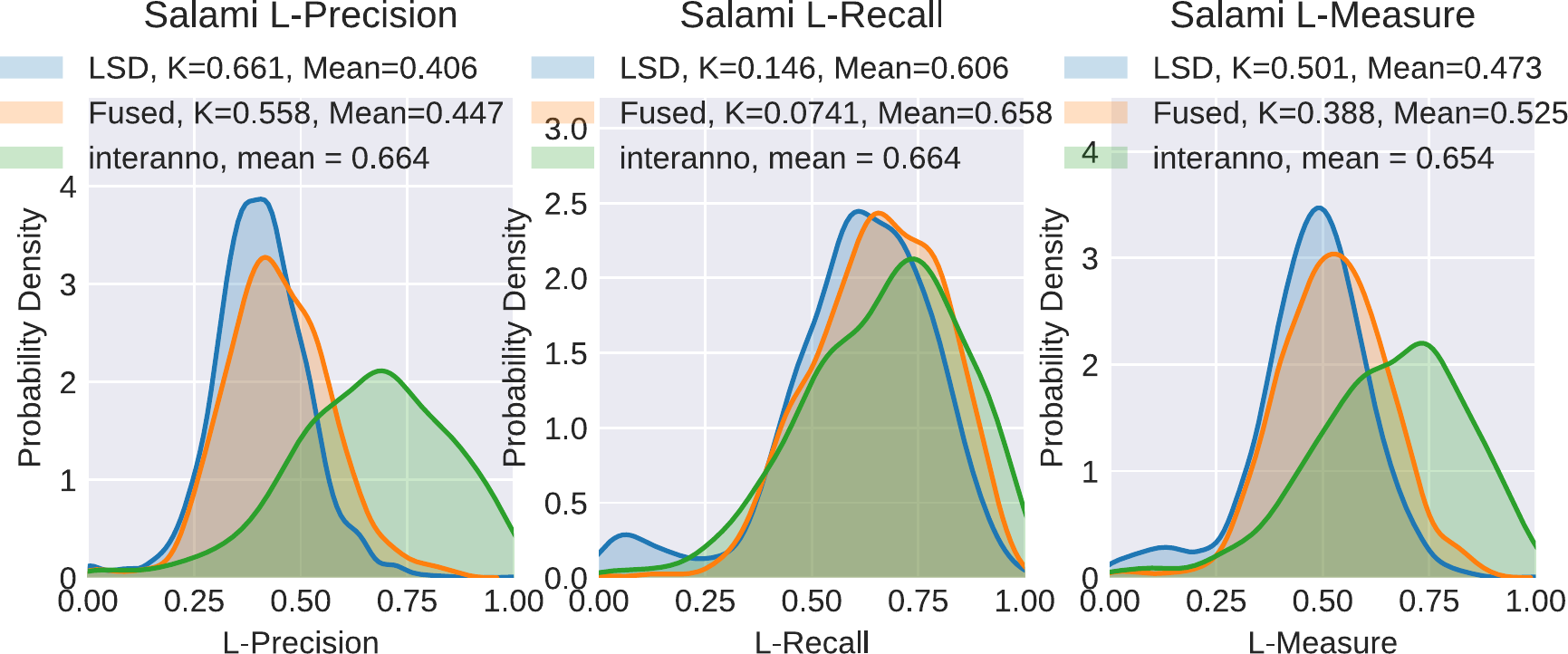}
    \caption{Distributions of L-precision, L-recall, and L-measure for inter-annotator agreement, the spectral clustering technique of~\cite{mcfee2014spectral}, and our fusion result.  The Komolgorov-Smirinov statistic and mean score are shown in the legends.}\label{fig:PRLDists}
\end{figure}

\begin{table}\label{tab:salami}
\footnotesize
\begin{tabular}{lrrrrrr}
    \toprule
{} &  $\mu$(P) &  $K$(P) &  $\mu$(R) &  $K$(R) &  $\mu$(L) &  $K$(L) \\
\midrule
Inter-Anno         &           0.664 &      ---      &        0.664 &       ---   &         0.654 &         ---\\
\midrule
MFCCs             &           0.371 &         0.663 &        0.295 &      0.617 &         0.283 &       0.713 \\
Chromas           &            0.320 &         0.767 &        0.287 &      0.717 &         0.271 &       0.792 \\
Tempogram         &           0.337 &         0.768 &        0.464 &      0.476 &         0.382 &       0.678 \\
CREMA             &           0.392 &         0.668 &        0.529 &      0.342 &         0.441 &       0.558 \\
Fused MFC/Chr &           0.422 &         0.601 &        0.612 &      0.163 &         0.491 &       0.465 \\
Fused Tgr/CRE &           0.388 &          0.670 &        0.631 &      0.119 &         0.473 &       0.501 \\
Fused $\kappa=3$             &           \textbf{0.447} &         \textbf{0.558} &        \textbf{0.658} &     \textbf{0.074} &         \textbf{0.525} &       \textbf{0.388} \\
Fused $\kappa=10$      &           0.424 &         0.606 &        0.623 &      0.167 &         0.498 &       0.445 \\

LSD\cite{mcfee2014spectral}          &           0.406 &         0.661 &        0.606 &      0.146 &         0.473 &       0.501 \\
\bottomrule
\end{tabular}
\caption{The means $\mu$ and $K$-scores of L-precision (P), L-recall (R), and L-measures (L) for different segmentations.}
\end{table}

Figure~\ref{fig:PRLDists} shows probability density functions for L-precision, L-recall, and L-measure for our technique with a hierarchy of 2--10 clusters and $\kappa=3$.  The distributions for our fusion are closer to human level agreement than those of LSD~\cite{mcfee2014spectral}, and they also correct a cluster of failure cases present in~\cite{mcfee2014spectral}.  Table~\ref{tab:salami} shows the mean (higher is better) and $K$-scores (lower is better) of precision, recall, and L-measure for individual features and various combinations of fusion (MFCC/Chroma, Tempogram/CREMA, and all).  In all cases, fusion improves over individual features, and fusing all features performs the best across all statistics.  We also show that using a smaller number of neighbors $\kappa$ for the spatial bandwidth is advantageous, as it tends to promote diagonal regions in the fused affinity matrices without connecting dissimilar blocks.

\section{Discussion / conclusion}

This work has shown promise of SNF + spectral clustering for hierarchical structure annotations, and we believe there will be other applications of feature-level SNF on affinity matrices in MIR.  There is also room for general theoretical development of the interplay between SNF and the graph Laplacian.

\newpage
\bibliographystyle{IEEEbib}
\bibliography{refs}

\end{document}